# Recommending and Release Planning of User-Driven Functionality Deletion for Mobile Apps

Maleknaz Nayebi, Konstantin Kuznetsov*†*, Andreas Zeller*†*, Guenther Ruhe

[1]EXINES Lab, York University, Toronto, Canada.
[2]Saarland University, Saarbrücken, Germany.
[3]CISPA Helmholtz Center for Information Security, Saarbrücken, Germany.
[4]SEDS Lab, University of Calgary, Calgary, Canada.

*Corresponding author(s). E-mail(s): mnayebi@yorku.ca;
Contributing authors: konst.kuznetsov@icloud.com; zeller@cispa.de;
ruhe@ucalgary.ca;
*†*These authors contributed equally to this work.

**Abstract**

Evolving software with an increasing number of features poses challenges in terms of comprehensibility and usability. Traditional software release planning has predominantly focused on orchestrating the addition of features, contributing to the growing complexity and maintenance demands of larger software systems. In mobile apps, an excess of functionality can significantly impact usability, maintainability, and resource consumption, necessitating a nuanced understanding of the applicability of the law of continuous growth to mobile apps.
Previous work showed that the deletion of functionality is common and sometimes driven by user reviews. For most users, the removal of features is associated with negative sentiments, prompts changes in usage patterns, and may even result in user churn. Motivated by these preliminary results, we propose Radiation to input user reviews and recommend if any functionality should be deleted from an app's User Interface (UI). We evaluate Radiation using historical data and surveying developers' opinions. From the analysis of 190,062 reviews from 115 randomly selected apps, we show that Radiation can recommend functionality deletion with an average F-Score of 74% and if sufficiently many negative user reviews suggest so. We conducted a survey involving 141 software developers to gain insights into the decision-making process and the level of planning for feature deletions. Our findings indicate that 77.3% of the participants often or always plan for such deletions. This underscores the importance of incorporating feature deletion planning into the overall release decision-making process.

# 1 Introduction

Lehman's laws of software evolution [1] highlight the importance of continuous adaptation to prevent a decline in user satisfaction over time. Lehman's sixth law extends this idea, stating that a program's functional content must continually grow to maintain user satisfaction throughout its lifespan. While this holds true for service-oriented platforms like operating systems, where maintaining functionality is crucial for backward compatibility, it poses a challenge for programs primarily used by individuals. In such cases, a constant increase in features conflicts with usability as more features compete for user attention. As Buschmann [2] pointed out, there is a risk of trading functional coverage for quality as the reliability, performance, and maintainability are postponed to the time "when the functionality is stabilized". The concept of excessive software development emerges as a recognized concern [3]. However, conventional release planning, often fixated on the addition of features, may inadvertently compromise quality in the pursuit of comprehensive functional coverage [4, 5].

Mobile apps, constrained by factors such as small screens and limited resources, exemplify the delicate balance between functionality and usability, where adding functionality comes at a cost [6]. In navigating these constraints, developers may find it advantageous to consider the removal of functionality that detrimentally influences the user experience [7], challenging the conventional notion of perpetual growth. This becomes especially crucial when considering user-centric principles, where ease of use and discoverability are key. Hence, developers should be interested in *removing* functionality that negatively impacts the user experience [7].

Developers of mobile apps face the challenge of optimizing the user experience by strategically deciding when to add or remove features [8, 9]. While this removal can be the result of different development activities (for example, removing the code, commenting out the code, or disabling respective UI elements), from the user's perspective, a functionality is considered removed when it is no longer accessible through the user interface [10, 11]. Empirical studies on mobile app release engineering highlight developers' increasing awareness of the impact of user feedback on code changes. However, techniques for release planning have not considered the removal of functionality.

There is an established body of knowledge on the release engineering of mobile apps. Several techniques [12] have been proposed for the release planning of mobile apps. Generally, these existing methods are focused on feedback development planning based on user reviews. They first categorize reviews into general categories of uninformative comments, feature requests, bug reports, or praise. Then, they aim to satisfy that user feedback in the upcoming release. The large number of user reviews on mobile app stores, which can range from zero to millions of reviews per release [13, 14] prompted several studies to summarize and prioritize user concerns for enhancing mobile apps [15–17].

Palomba et al. [18, 19] confirmed empirically that mobile app developers are changing their code based on the crowdsourced app reviews. Among these studies, multiple provided a variety of taxonomies for mobile app reviews [20, 21]. When analyzing user reviews, a few studies reported a reason for negative reviews [22, 23]. Further, in our previous study [7], we analyzed commit messages of mobile apps and established a taxonomy of "what", "how" and "why" deletions occur in code repositories. That is, these deletions range from the code deleted from a repository with no specified reason (e.g., accidental removal) to the code updated during refactoring to improve the code structure and UI elements removed due to undocumented reasons. In this context, and with the emergence of data-driven decision-making, machine learning has also impacted how organizations approach release planning. As the software development landscape continues to evolve, we can now extend release decisions to encompass a broader domain and feature deletions by leveraging historical data and predictive analytics.

This paper is an extension of the study published and awarded in a conference [11] and was invited for this journal extension. In this version, while summarizing some of the findings in the first research question, we further :

- Enhance the motivation for our research by presenting a thorough literature review on feature deletion. Integrate our user study findings into the background section,
- Provide extended information on the user study (RQ2),
- Provide a survey with developers on the possibility and support of release decisions and documenting current best practices.

In our paper, we follow a clear structure based on the design science process [24, 25]. We start with problem conceptualization, outlining the problem we're addressing and connecting the evidence from theory and practice. Then, we move on to designing solutions that we believe can recommend functionality deletion by offering Radiation. Finally, we validate the empirical solution we offered following the design science process. We also reflected upon the problem as an important element for this knowledge transmission by surveying the possibility and importance of including feature deletions in the release planning. This structured approach helps us ensure that our work is thorough and practical, leading to meaningful insights and solutions.

In what follows, in Section 2, we explain some background knowledge needed to make this paper self-contained and motivate the added research question to this study. We then move to problem conceptualization in Section 3. We present our solution design in Section 4 and discuss protocols to validate it from multiple aspects in Section 5. We then present the results of this validation in Section 6. We then present the results of a survey with developers to discuss the current status quo for considering functionality deletions in the release decisions 7 and move to present the threats to the validity of our study in Section 8. We then present the related work in Section [9](). We wrap the paper by discussing the future work (Section [10]()) and the conclusions (Section [11]()).

## 2 Background: Software Functionality Deletions

In a mining study in 2018, we took a step to look into the code changes and investigate the evolution of open-source Android mobile apps [7]. We aimed to understand the frequency and nature of size reduction in releases, aiming to motivate the analysis of functionality deletion. The study involved 1,519 apps and over 20,806 GitHub releases. We compared the size of the code base (which was highly correlated with the size of APK file, 0.86). Our analysis showed that 98.8% of apps decreased their size at least once, with 61.3% experiencing more than a 10% reduction in size in at least one release.

We also analyzed the number of Android components (Activities, Services, Content Providers, Broadcast Receivers) as a proxy for functionality deletion. Notably, 37.6% of apps had decreased activities, and various changes in services, providers, and receivers were observed across releases. We took a step further and used the Backstage tool [26] to examine the deletion of UI elements and associated API calls in a subset of apps. The results indicated that 39.8% of apps had UI elements removed in at least one release. Our findings show that nearly one-third of the scrutinized apps exhibit a discernible decline in size, activities, and UI elements over successive releases. This finding reinforces our initial hypothesis, suggesting that the conventional notion of continuous growth in functionality, as posited by Lehman's laws, calls for an investigation into the dynamics surrounding functionality deletion in software evolution.

We further reported on the in-depth analysis of 8, 000 commit messages from these apps to understand the "What," "Why", and "How" of functionality deletions. In terms of *what functionality was deleted,* we identified a total of 22 categories of functionality deletion. These categories were subsequently organized into a two-level taxonomy featuring four high-level categories: "security and privacy elements" (such as licensing or permissions), "communication bridges", "user interface elements", and "development artifacts".

In exploring *why functionality was deleted,* we presented a taxonomy with 13 categories, grouped into broader themes as "improving user experience", "improving the quality of the existing code", "Better use of resources", and "Better communication". We also identified an "Unknown" category for commits lacking a description of the reason for deletion.

The retrospective analysis revealed that 29.98% of functionality deletions are related to UI elements, and 11.27% of functionality deletions are intended to improve users' experience. Conducting a thorough analysis of commit messages, we provided an in-depth exploration of functionality deletion in mobile apps. Notably, 11.23% of commits cite the enhancement of user experience as the rationale for deletion. Additionally, the author's examination of commit messages reveals that 14.63% of deletions are influenced by negative user feedback. It is worth noting that, despite these findings, there has been no empirical evaluation of users' perceptions regarding feature removal.

Furthermore, the app developers emphasized that multiple factors impact decisions about functionality deletions. Complexity and required maintenance effort, extent of usage, and user reviews with specific attributes were identified as the top three most important factors. Among the characteristics highlighted were annoyed reviews, reviews expressing similar concerns about the app, and reviews associated with low ratings.

## 3 Problem Conceptualization: Importance of Feature Deletions to Users

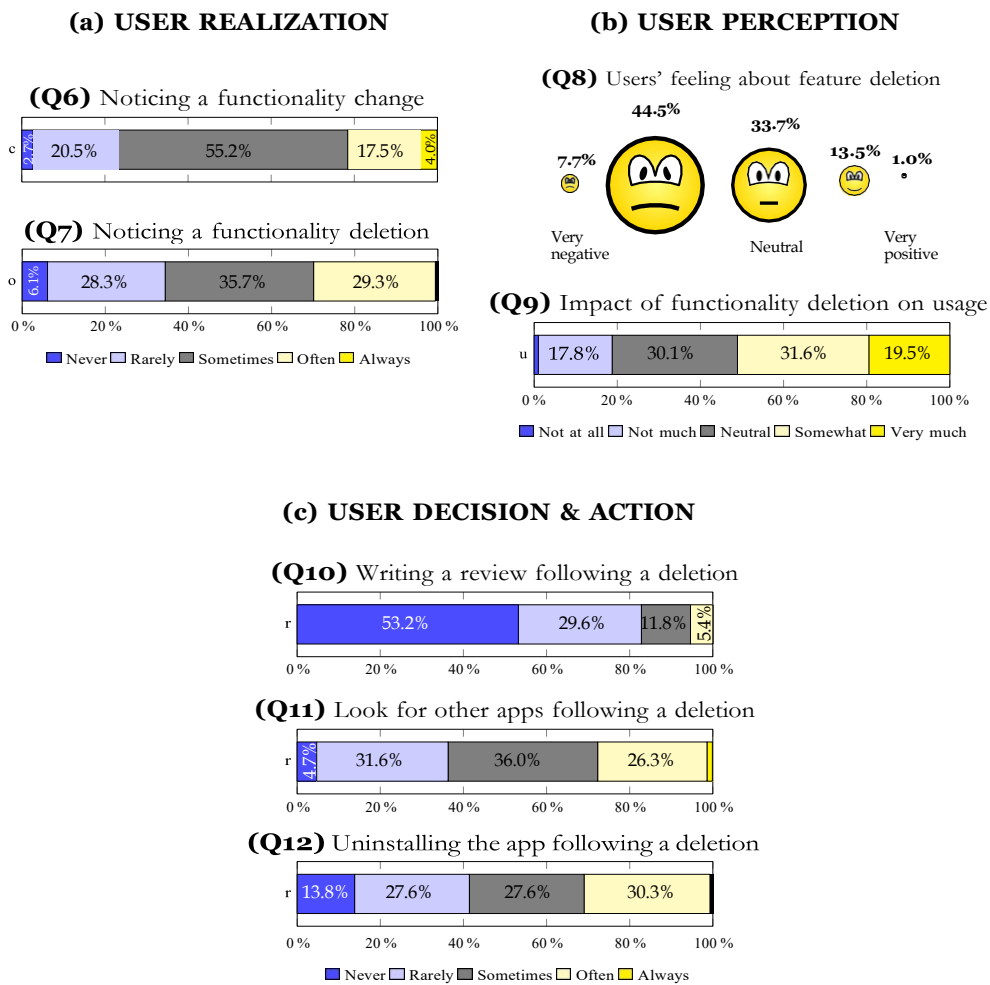

Fig. 1: Results of the survey with app users (Q6 to Q12).

To conceptualize the problem, we conducted an in-depth survey with users, which we detailed in the conference version of this study, aiming to investigate the significance of feature deletion for mobile app end-users and answer the question: "How do mobile app end-users perceive the deletion of software functionality?". Following established guidelines for survey research [27], our study comprised four primary components. Our survey began by collecting participants' demographic information. Subsequently, we explored participants' awareness of missing features or functionalities across different app releases. Following this, we assessed whether the deletion of features influences users' overall satisfaction with the mobile app. Lastly, we inquired about the extent and impact of functionality deletion or limitation on participants' app usage patterns. The survey comprised 12 closed-ended questions, with five initial questions dedicated to gathering demographic information. The remaining queries utilized a five-point Likert scale to gauge participants' opinions.

297 individuals completed our survey [11]. Among these participants, 44.1% fell within the 28-40 age range, 27.3% were between 18-28 years old, 15.5% were in the 40-64 age bracket, and 13.1% were above 64 years old. Regarding app installation, a majority (51.9%) reported having personally installed 5-10 apps on their devices, while 26.9% installed more than ten apps, and 21.2% installed fewer than five apps. Regarding daily app usage, 53.9% used more than ten apps daily, with only 1.3% using less than five apps daily and 44.8% using 5-10 apps daily. Among the respondents, 80.1% had uninstalled some apps, but only 39% occasionally or more frequently left reviews for mobile apps.

Figure 1 highlights the main findings of our survey. According to our survey, a majority of users (55.2%) reported *sometimes* noticing changes in the features of the mobile apps they use. Regarding feature deletions, 34.4% *never* or *rarely* noticed deletions, while 65.7% reported sometimes or more frequent awareness of feature deletions. These indicate the extent to which participating users realize and notice the change and deletion in mobile app features.

As for the perception of users toward a feature deletion in an app and its impact on their app usage, we found that approximately 51.9% of participants expressed a somewhat *negative* sentiment associated with feature deletions, with 41.1% stating negative and 7.75% stating *very negative* sentiments. Conversely, 13.5% had a *positive* perception, and 1.0% expressed *very positive* feelings about feature deletions. Regarding the impact on app usage, 48.8% reported *no change,* while 51.2% reported *somewhat* or *extensive* changes in app usage following a feature deletion.

When asking the extent to which deletions impact users' decisions and provoke a reaction, only 17.17% of participants *often* or *sometimes* left a review for a mobile app after a feature was deleted. In response to losing access to app functionality, 63.7% of participants *sometimes* or *more frequently* sought alternative apps. Additionally, 31% of participants reported *at least once* uninstalling an app due to a feature deletion, while 41.4% *never* or *rarely* did so, and 27.6% *sometimes* took this action. Deletion of app functionality provokes negative feelings for the majority of the participants (51.9% of the participants) and somewhat changes their usage behavior (51.2% of the participants). Functionality deletion caused 31.0% of the users often to migrate to another app. 27.6% of the users uninstalled the app following the deletion of a feature.

# 4 Solution Design

The results of our survey with end users motivated us to further evaluate the significance of feature deletions. We are interested in studying the feature deletion within the evolution process and release planning of mobile applications from user perspective. As the functionality is usually exposed to the user via (G)UI elements [28], in this study, we are particularly interested in the deletions visible to the end user. App reviews are categorized around these UI elements [19].

To assist the production team with such decisions, we introduce Radiation[1] to recommend deletions based on user reviews. We further evaluate Radiation's performance retrospectively and by performing cross-validation. To externally validate Radiation, we surveyed 37 software developers and 42 users to understand their perception of the value of deletions recommended by Radiation. Radiation predicts and recommends functionality deletions in mobile apps.

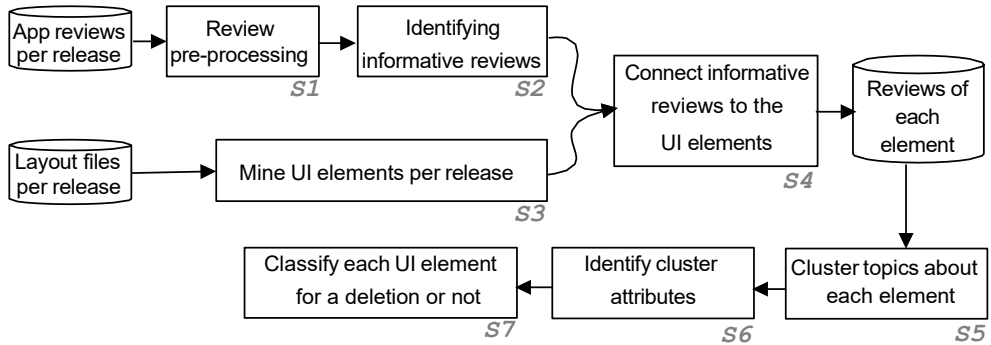

**Fig. 2**: The process of Radiation to support decisions on user-driven UI functionality deletions.

Multiple factors may trigger functionality deletion. We designed Radiation to recommend deleting functionalities suggested by user reviews. The current literature on apps' user needs and planning is primarily focused on adding features or fixing bugs in each release, based on user requests [20, 29, 30]. Radiation differs from this approach by targeting deletions and inputting user reviews. Radiation is a recommendation tool that helps developers identify deletion candidates. While deleting features is sometimes necessary [7], developers must be cautious about the features they removed, as it can result in a negative user experience and potentially losing customers, as shown by our survey study (see Section 2). Radiation is the first step to assist developers with this task. Figure 2 illustrates the six steps of Radiation. We relied on the best results presented in the literature to design each step.

*Step 1.* **Reviews pre-processing.** (following [31]) We eliminated emojis, special characters, and stop words and expanded contractions ("can't" was expanded to "can not"). Then, we applied lemmatization to map the words into their dictionary format ("deciding" and "decided" turned into "decide"). We used Python library NLTK for this step. We customized the list of stop words as suggested by Maalej and Nabil [32] and Palomba et al. [18].

---

[1]Radiation = **R**eview b**A**sed **D**elet**I**on recommend**ATION**

*Step 2.* **Separating informative and non-informative reviews.** (following [31]) Not all reviews were useful. We followed the definition of what is informative and non-informative as described by Maalej and Nabil [32]. In short, informative reviews communicate content that can be used in the process of the app evolution, while an advertisement, a short statement of praise (i.e., "The app is nice"), or a statement of an emotion (i.e, "I hate this app!") is not informative for enhancing an app in future releases. To identify informative reviews, we manually classified a fraction of reviews and used them to train a Naive Bayes classifier (following [32]).

**Table 1**: Features used in RF to recommend if some functionality is a candidate for deletion.

| Attribute | Reason | Description |
|---|---|---|
| $\|Reviews\|$ | [29, 31] | The number of reviews in a cluster. |
| $\overline{rating}$ | [29, 31] | Each app reviews is associated with a rating. $\overline{rating}$ is the average rating of reviews in a cluster. |
| $\Delta\ rating$ | [29] | $\Delta$ between the average rating of the cluster and the average rating of the app in a specific release. |
| $\overline{polarity}$ | [20, 30, 33] | The average polarity of the reviews in a cluster. Polarity is one dimension of sentiment and is a number between $[-1, 1]$. $-1$ shows negative sentiment, 0 is neutrality, and 1 is the very positive feeling. |
| $\overline{objectivity}$ | [7] & our user survey | Average objectivity of the reviews in a cluster. Objectivity is another aspect of sentiment and is a number between $[0, 1]$. 0 shows the message is totally objective (expression of facts) and 1 shows the message was opinionated (subjective) [34]. |
| $\|uninstall\|$ | [7] & our user survey | The number of reviews talking about "uninstalling the app or requesting "refund". |

*Step 3.* **Finding UI elements for each release.** (following [29]) For *each release,* we extracted UI elements used in an application. We leveraged the UI elements to connect the reviews with the apps' functionality following the method of Palomba et al. [18]. They showed that users write reviews related to the app components visible to them, which are the elements of the user interface. To mine UI elements, we implemented the lightweight analysis of Android layout files. These files include most of the GUI elements, also known as view widgets, and control as it is visible to the app user [26, 35]. Additionally, we parsed the Strings.xml file, which contains text strings for an app. By mining these files, for each identified UI element, we got its *description* consisting of an element type, a variable name used in the code, a label associated with the element, and an icon name if applicable (e.g., <Button, btn_mic, 'Start Listening',_>).

*Step 4.* **Connecting reviews to the UI elements.** (following [18]) We used the description of elements connecting reviews to app functionalities. To connect a review to a UI element in a release $V_i$, we calculated the cosine similarity between the text of a UI description and a review's content. We established a connection when the similarity score exceeded a threshold of 0.65. Palomba et al. [18] used the threshold of 0.6 for this purpose. However, when analyzed manually, we slightly increased the threshold to achieve a more accurate matching.

*Step 5.* **Clustering reviews based on their topic.** (following [18]) Several app reviews are pointing to the same functionality, while they may contain different opinions about that functionality. We used *Hierarchical Dirichlet Process* (HDP) [36] with its default setup to group reviews related to each functionality (UI element) as suggested by Palomba et al.[37]. HDP is a topic-mining technique that automatically infers the number of topics and is an extension of LDA [38], which is designed based on a non-parametric Bayesian network [39]. The difference with LDA is that the number of topics does not need to be specified in advance for HDP. Using HDP as described in [37], we performed topic modeling and formed clusters with reviews about a particular topic associated with a UI element. In this way, every cluster represents a set of similar concerns about a UI element. One review might also discuss multiple UI elements; hence, the clusters are non-exclusive. We manually analyzed the results for 1,500 reviews across eight apps: The topics were intuitive and understandable.

*Step 6.* **Identifying candidate functionality deletion.** Following the existing literature on prioritizing app reviews (Table 1), we selected attributes for identifying and recommending possible functionality deletion. To determine candidates, we used Random Forest, as suggested by related studies [29], and showed good time performance. A list of attributes for training is presented in Table 1. The "polarity" and "objectivity" of the reviews in a cluster were extracted by sentiment analysis performed by Pattern [13, 34] technique.

Radiation involves six main steps. We used established and top-performing state-of-the-art methods in forming our Step 1, Step 2, and Step 3. Radiation also adheres to state-of-the-art approaches for clustering user reviews around UI elements (Step 4). It has been established [19] that user reviews often address UI elements as the app functionalities visible to the end user. In particular, Radiation is designed to predict and recommend the deletion of UI functionality based on user reviews. It is important to note that our intent was to demonstrate whether recommending feature deletion is possible rather than implementing the most performant methods. We evaluated the results of Radiation in two ways: first, by retrospectively comparing the decisions that were actually made in the app, and second, by externally evaluating with app developers to understand their perception of the Radiation recommendation (external evaluation). We explain these two in the next section.

# 5 Empirical Validation of Radiation

In this section, our objective is to internally evaluate the performance of Radiation and, in particular, evaluated *how effectively can functionality deletion be recommended* *based on user reviews?*. We address this question through two approaches: firstly, retrospectively comparing automated recommendations with actual deletions, and secondly, conducting an external evaluation. For the latter, we engaged software developers to assess whether a functionality should be deleted from an app based on provided reviews, then compared their decisions with the outcomes of our approach. When validating our solution approach, we answer three research questions:

***RQ1:*** How effectively can functionality deletion be recommended based on user reviews? For a set of 190,062 reviews, we applied Radiation to identify the reviews that provoked the deletion of functionality. We evaluated Radiation internally (via cross-validation) and externally (with 37 developers):

**RQ1-1** How does the recommendations compare with actual deletions?

**RQ1-2** To what extent do app developers consider analogical reasoning useful for predicting functionality deletions?

***RQ2:*** **What is users' experience with the functionalities that *Radiation* offers for deletion?**

We conducted a survey with 42 participants who used the app in the lab to assess their sentiment towards the functionalities recommended for deletion by Radiation. After familiarizing themselves with the app, we asked each participant to evaluate 30 UI functionalities based on their level of liking and the importance of deletion. We performed a controlled experiment by presenting the question for the features recommended for deletion by Radiation and those not recommended for deletion. Finally, we analyzed the relationship between user sentiment and the recommendations provided by the tool. The end-user study confirmed the recommendations' validity.

## 5.1 Protocols for Internal Validation of Radiation Recommendations with Developers (RQ1-1)

For this internal and retrospective evaluation in **RQ1**, we excluded apps with fewer than two releases (554 apps) and randomly selected 10% (115 apps) from the remaining 1,150 for in-depth analysis with an overall of 3,364 releases. These 115 apps encompassed an overall 190,062 reviews and had an average of 176 reviews per month.

To validate the effectiveness of Radiation, we retrospectively compared its recommendations with actual source code changes across 115 apps and 3,364 releases. In Step 5 of the Radiation process, we clustered reviews for each UI element and labeled clusters as "deleted" or "not deleted" through manual inspection of source code commits [7]. Two annotators performed this labeling, through which we achieved a 96% Kappa agreement rate, and discrepancies were promptly resolved through a brief code look-up and rechecked by the first author of the paper as the moderator. As such, and to create a *truth set*, we tagged reviews in $V_{i-1}$ as "deleted" if the corresponding element $E_i$ was deleted in release $V_i$. Subsequently, we internally validated Radiation by comparing its predictions with this truth set. Hence, if an element $E_i$ was deleted in release $V_i$, we tagged the clustered reviews in $V_{i-1}$ as "deleted". As such, each cluster is a set of reviews with similar criticism relevant to a UI element. We used these manually labeled clusters as our *truth set*. To internally validate our results, we compared the output of Radiation with this truth set. When comparing the results of Radiation with the code changes retrospectively, one of the four outcomes was observed:

*TP:* Radiation recommends deletion of $E_i$ in $V_i$, and historical data of our truth set shows the element was deleted.
*TN:* Radiation does not recommend deletion of $E_i$ in $V_i$, and historical data of our truth set shows the element was not deleted.
*FP:* Radiation recommends deleting $E_i$ in $V_i$, but our truth set's historical data shows that the element was not deleted.
*FN:* Radiation does not recommend deletion of $E_i$ in $V_i$, but historical data of our truth set shows its deletion.

This retrospective analysis resulted in a confusion matrix, enabling the calculation of precision, recall, and F-Score for Radiation.

## 5.2 Protocols for External Validation of RADIATION with software developers (RQ1-2)

To assess the external validity of Radiation recommendations, we conducted analysis with software developers of the apps. Developers were given the cluster of user reviews. Each of these clusters consisted of reviews about a UI element (generated in Step 5 of Radiation). Given the cluster of user reviews, we asked each developer, "Based on your understanding of the given reviews, please categorize each cluster as either motivating functionality deletion or not motivating functionality deletion." Subsequently, we compared these developers' judgments with the Radiation's outcomes, introducing the possibilities of true positives (TP), true negatives (TN), false positives (FP), and false negatives (FN). This evaluation relies on developers' subjective assessments rather than historical data and differs from the previous section's process. To mitigate bias, we randomly selected 25 apps, and three developers independently evaluated each functionality cluster, with final decisions determined by majority agreement.

Our objective was to gauge software developers' perspectives on the accuracy of Radiation recommendations. Initially, we invited developers who contributed to repositories from our set of F-Droid open-source apps. However, due to limited availability and responsiveness, we also perform recruitment through social media and professional networks. Through non-compensated and convenience sampling, 37 developers with an average of 8.3 years (ranging from two to 15 years) of overall software development experience and 4.4 years of mobile app development experience (ranging from one to 12 years) were enlisted. Each developer contributed to the development of at least two apps. In conducting this evaluation, the developers reviewed recommendations for 25 apps and analyzed 36,039 reviews, constituting an assessment for 20% of our selected apps for validation.

Further, the quality of topics and modeling in Step 5 is crucial to the success of Radiation. To assess the effectiveness of clustering by HDP in Step 5 of Radiation, we utilized a human judgment method called *topic intrusion* [40]. This involved presenting the top two topics with the highest similarity for a review and presenting them along with a random topic of lower probability (the intruder topic) to a developer, who was then asked to identify all relevant topics.

**LOOP HABIT TRACKER**

**Q:** Do you need more time to familiarize with the app?  ○ Yes  ● No

**Q:** Please rate your overall expereince with the "LOOP HABIT TRACKER" app.  ★★★☆☆

**Q:** In your opinion, how important is it to have "Detailed Scoring of Daily Progress" feature as part of this app?

○ It is essential  ● It is worthwhile  ○ it is unimportant  ○ it is unwise

**Q:** How did you like the "Detailed Scoring of Daily Progress" feature?

○ Strongly Disliked (-2)  ● Disliked (-1)  ○ Neutral (0)  ○ Liked (+1)  ○ Strongly liked (+2)

**Q:** How do you feel if the "Detailed Scoring of Daily Progress" feature is being deleted?

○ Strongly Disliked (-2)  ○ Disliked (-1)  ○ Neutral (0)  ● Liked (+1)  ○ Strongly liked (+2)

**Fig. 3**: Questions asked for evaluating Radiation with users for a sample app "Loop habit tracker" in **RQ2**.

37 developers participated in our study and evaluated clustering for 36,039 reviews. To evaluate the results of Step 5, we calculated *Topic Log Odds (TLO)* [40]. TLO is a quantitative measure of agreement between a model and a human. TLO is defined as the difference between the log probability assigned to the intruder topic and the log probability assigned to the topic chosen by a developer. This number is averaged across developers to get a TLO score for a single document $d$ [41]:

$$TLO(d) = \frac{\sum_{s}^{L} \log \theta_{r,trueintruder} - \log \theta_{r,intruderselectby's'}}{S}$$

Where $\theta_{r,t}$ is the probability that a review $r$ belongs to a topic $t$, and $S$ is the total number of developers.

## 5.3 Protocols for external validation of RADIATION with users

We installed the app on our own devices. We recruited participants from our personal network by advertising on social media and mailing lists following a convenient sampling protocol [42].

On the day of the study, we arranged the devices with the installed app for the participants, ensuring they were fully charged and functioning correctly. We also provided a designated area for participants to engage with the app comfortably. We provided a consent form outlining the study's purpose, procedures, and participant rights should be prepared for participants. We further provided a brief overview of the study and explained the purpose, emphasize the voluntary nature of participation, and obtain informed consent.

We deliberately involved only one participant for each study session to mitigate the influence and peer pressure. We provided tablets and instructed participants to actively engage with the app for at least 20 minutes, exploring various features and sharing feedback on their experience. Following this period, we checked in with participants to see whether they required additional time or were prepared to move forward. Following this interaction phase, we included questions about their overall experience.

We provided participants with a structured questionnaire asking them to rate their experience with the app, similar to the app store, on a one- to five-star scale. We then asked them to provide feedback on a particular app feature.

We asked the participants to rate the importance of a given feature following a Kano model [43–45] either as essential, worthwhile, unimportant, or unwise.

Further, we introduced a hypothetical scenario where a specific app feature is removed. We asked participants to provide feedback on their feelings and the potential impact on their overall experience. Figure 3 shows a sample of this task. We randomly selected 30 UI elements and functionalities from each app as part of our evaluation. We made a deliberate effort to include a mix of correct (TP and TN) and incorrect (FP and FN) deletion recommendations (as explained in RQ2), whenever possible. In total, we evaluated 650 UI functionalities, with 325 recommended for deletion by Radiation and 325 that were not recommended for deletion. Our survey included 42 participants selected via non-compensated and convenience sampling from our social and professional network. For each functionality of the app, three users provided evaluations. Figure 3 displays a sample survey question and the response of one participant specifically for the org.isoron.uhabits app.

After familiarizing themselves with their assigned apps for at least 20 minutes, we presented a specific feature of the app they had studied. In this context, we consider the participants in our evaluation as "users" and will refer to them as such, noting their controlled level of experience with the app.

Then, we requested that they rate their liking of the feature on a five-point Likert scale. Furthermore, we also asked the participants to express their emotions if the feature were to be removed. We used conventional sentiment scores [46] for evaluation, with −2 indicating strong dislike, 0 indicating neutrality, and +2 indicating strong liking.

We gathered the data only through the formal questionnaire provided to the participants and the questions outlined above.

## 6 Validation Results

We used open-source Android apps for this evaluation. As of June 2022, F-Droid (the open-source repository for Android mobile apps) included 3,810 mobile apps. We identified 1,704 apps with a valid link to their GitHub repositories. These apps involve a total of 14,493 releases. As deletions are identified by comparing sequential releases, deletions are only meaningful if the app has at least two releases. So, we excluded 554 apps with fewer than two releases from our analysis to evaluate Radiation over multiple releases. We gathered reviews from the Google Play store for the remaining apps while accessing their code and development artifacts through GitHub.

We randomly selected 8,300 reviews (≅ 5% of the total number of reviews) across different apps and manually labeled each review as "informative" or "non-informative" as described in Step 2 of Radiation. We followed the definition of informative and non-informative as described by Maalej and Nabil [32]. Two researchers classified these reviews with an average Cohen's Kappa agreement degree [47] of 86%. We labeled 2,917 of these reviews as "non-informative" and used them along with the same number of "informative" reviews randomly sampled from the rest of the reviews to train a classifier. Finally, we identified 8.1% of the total number of reviews as uninformative. When it came to the performance of the Naive Bayes classifier used for the automatic separation of these reviews in Step 2, we achieved an F1 score (the harmonic mean of precision and recall [48]) of 0.82. This score was calculated as the average of ten 10-fold cross-validation runs. We created recommendations using Radiation and analyzed 115 randomly selected apps in detail. We then used these recommendations as well as evaluated Radiation against developers' judgment (**RQ1**) and users' experience (**RQ2**) for 25 apps. When we performed internal and external validation of our method.

## 6.1 Results of internal validation of solution approach

Table 2 presents the results of **RQ1** and **RQ2** for 25 apps that were cross-validated and evaluated by developers. Figure 4 demonstrates the goodness of the topic modeling of app reviews (Step 5) as part of **RQ2**).

We conducted cross-validation on 115 apps, 3,364 releases, and a total of 190,062 reviews. 8.1% of this total number of reviews were uninformative. The results indicate high precision (*0.83*) and recall of *0.48* using 10-fold cross-validation. The precision is considerably higher than recall because in Radiation, the number of false positives (FP) is much lower than false negatives (FN). In other words, in mobile apps, there have been deleted features, but Radiation cannot recommend them for deletion (FN). This results in a low recall. Radiation cannot (and is not designed to) capture all deletions that happen within a mobile app. However, as the first study looked into functionality deletion, we could predict with 83% precision. For several of these "false negatives", we did not find reviews related to an element that has been deleted. Hence, we concluded that the feature would not be deleted, and there were other reasons than user reviews for deleting the UI element. Table 2 details the confusion matrix for the 25 apps that were also externally evaluated in **RQ2**. As the result, Radiation demonstrates 83% precision in recommending deletions based on user reviews. The low recall indicates that not all deletions in a mobile app are motivated by user reviews, which Radiation is not designed to capture.

## 6.2 Results of external evaluation with software developers (RQ2)

37 developers evaluated Radiation in two ways. First, by evaluating the quality of the topics created from reviews and about each UI as a result of Steps 4 and 5. Second, by assessing whether, as professional software developers, they would make the same decisions as Radiation regarding the deletion functionality based on user reviews.

**Evaluation of cluster topics about each UI element** We followed the approach of Palomba et al. [37] to cluster user reviews by their connection to UI elements. Hence, in Radiation, we first connected reviews to the UI elements (Step

**Table 2**: Evaluating results by comparing Radiation recommendations with (i) retrospective analysis of actual deletions and (ii) developers' perception. One user review might be relevant to multiple elements.

| App's package name | # of UI element across releases | # of reviews | Actual deletions (RQ1) | | | | | Developers' perception (RQ2) | | | | |
|---|---|---|---|---|---|---|---|---|---|---|---|---|
| | | | # of FP | # of FN | # of TP | # of TN | F1 score | # of FP | # of FN | # of TP | # of TN | F1 score |
| (A1) app.openconnect | 235 | 232 | 0 | 2 | 1 | 232 | 0.5 | 0 | 0 | 1 | 234 | 1 |
| (A2) com.google.android.stardroid | 1603 | 4480 | 0 | 2 | 18 | 1583 | 0.95 | 1 | 2 | 18 | 1582 | 0.92 |
| (A3) com.moez.QKSMS | 3009 | 2751 | 0 | 11 | 5 | 2993 | 0.48 | 2 | 4 | 5 | 2998 | 0.62 |
| (A4) com.vuze.android.remote | 774 | 494 | 0 | 2 | 8 | 764 | 0.89 | 1 | 0 | 7 | 766 | 0.93 |
| (A5) net.nurik.roman.muzei | 1088 | 4481 | 0 | 15 | 36 | 1037 | 0.83 | 1 | 0 | 35 | 1052 | 0.99 |
| (A6) org.androisoft.app.permision | 189 | 397 | 0 | 1 | 2 | 186 | 0.8 | 0 | 1 | 2 | 186 | 0.8 |
| (A7) org.connectbot | 471 | 4493 | 0 | 6 | 8 | 457 | 0.73 | 0 | 0 | 8 | 463 | 1 |
| (A8) org.dmfs.tasks | 862 | 207 | 0 | 7 | 7 | 848 | 0.67 | 0 | 4 | 7 | 851 | 0.78 |
| (A9) org.evilsoft.pathfnder.rference | 652 | 1520 | 0 | 0 | 2 | 650 | 1 | 1 | 0 | 1 | 650 | 0.67 |
| (A10) org.isoron.uhabits | 895 | 1976 | 3 | 31 | 101 | 760 | 0.86 | 4 | 13 | 100 | 778 | 0.92 |
| (A11) com.spazedog.mounts2sd | 394 | 497 | 3 | 7 | 60 | 324 | 0.92 | 2 | 0 | 61 | 331 | 0.98 |
| (A12) org.telegram.messenger | 840 | 73682 | 2 | 30 | 26 | 782 | 0.62 | 3 | 3 | 25 | 809 | 0.89 |
| (A13) in.blogspot.anselbros.torchie | 134 | 473 | 8 | 12 | 72 | 42 | 0.88 | 5 | 1 | 75 | 53 | 0.96 |
| (A14) com.emaguy.cleanstatusbar | 86 | 392 | 1 | 7 | 8 | 70 | 0.67 | 0 | 0 | 9 | 77 | 1 |
| (A15) com.boardgamegeek | 1317 | 506 | 33 | 224 | 191 | 6682 | 0.6 | 3 | 25 | 221 | 6881 | 0.94 |
| (A16) com.gelakinetic.mtgfam | 4510 | 2366 | 1 | 3 | 4 | 4502 | 0.67 | 0 | 1 | 5 | 4504 | 0.91 |
| (A17) org.addhen.smssync | 235 | 41 | 6 | 0 | 22 | 207 | 0.88 | 7 | 2 | 21 | 205 | 0.82 |
| (A18) com.amaze.filemanager | 620 | 1241 | 7 | 12 | 25 | 576 | 0.72 | 0 | 1 | 33 | 586 | 0.98 |
| (A19) com.gh4a | 344 | 301 | 4 | 8 | 14 | 318 | 0.7 | 1 | 1 | 17 | 325 | 0.94 |
| (A20) org.kontalk | 54 | 39 | 2 | 2 | 7 | 43 | 0.78 | 2 | 1 | 7 | 44 | 0.82 |
| (A21) org.transdroid.lite | 942 | 538 | 2 | 3 | 7 | 930 | 0.74 | 0 | 0 | 9 | 933 | 1 |
| (A22) de.qspool.clementineremote | 444 | 355 | 4 | 9 | 13 | 418 | 0.67 | 2 | 3 | 15 | 424 | 0.86 |
| (A23) com.daiancorp.mh4udtabase | 3101 | 979 | 29 | 51 | 73 | 2948 | 0.65 | 12 | 5 | 90 | 2994 | 0.91 |
| (A24) org.servalproject | 547 | 252 | 4 | 14 | 10 | 519 | 0.53 | 2 | 3 | 15 | 527 | 0.85 |
| (A25) org.wikipedia | 17830 | 15531 | 23 | 0 | 94 | 17713 | 0.89 | 1 | 0 | 116 | 17713 | 0.99 |
| **Average** | 1647.04 | 4728.96 | 5.28 | 18.36 | 32.56 | 1823.36 | 0.74 | 2 | 2.8 | 36 | 1838.84 | 0.9 |

**FP** (False-Positive): Recommended as deletion but was not, **FN** (False-Negative): Recommended not a deletion but it is, **TP** (True-Positive): Recommended as deletion and it is, **TN (True-Negative)**: Recommended as not a deletion and is not.

4) and then clustered the reviews around each UI element using HDP topic modeling (Step 5) [37]. Topic modeling was used as many users stated similar concerns in reviews, and each review might have contained multiple concerns about functionality.

We presented the number of UI elements along with the number of clusters and number of user reviews in Table 2. To evaluate the usefulness of our topic model, we relied on the judgment of app developers. After asking them to evaluate the topics using topic intrusion, we calculated TLO as suggested by Chang et al. [40]. We present the distribution of TLO in the boxplot chart of Figure 4. *TLO* = 0 shows the highest conformance between developers and the topic modeling technique. Comparison of the distribution of our HDP clustering showed a slight disagreement between developers and machine learning results as the median is around −3. However, this is still considered as a relatively low disagreement compared to former benchmarks [40, 41].

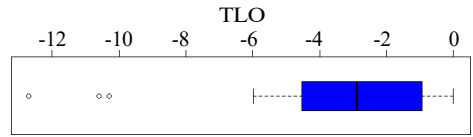

**Fig. 4**: Topic Log Odds (TLO) shows the performance of Radiation's clustering against developers' perception.

**Evaluating Radiation recommendations:** We asked developers to evaluate whether a cluster of reviews for a UI element was "motivating a functionality deletion" or "not motivating a functionality deletion" (e.g., implying a bug fix). We compared Radiation results to developer perceptions for 25 randomly selected apps, resulting in an average F-Score of 90% for Radiation. See Table 2 for the number of true and false recommendations for these apps.

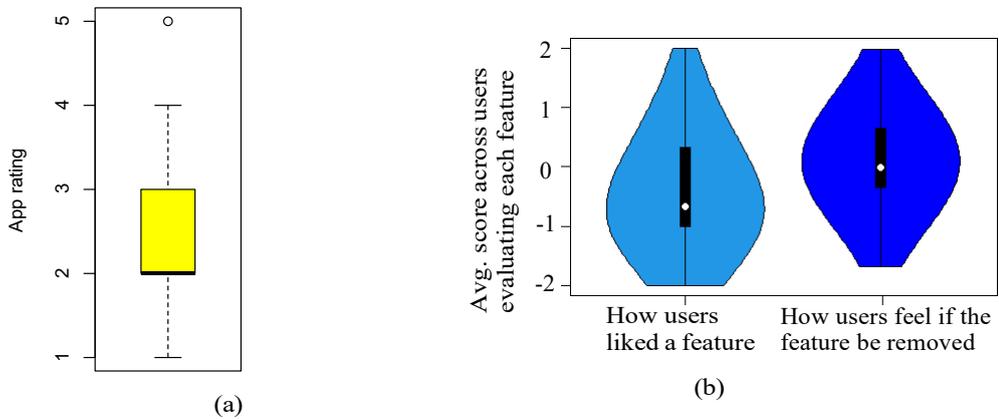

**Fig. 5**: (a) Overall app rating in the evaluated set, and (b) the Users' sentiment when evaluating 650 features (in blue) with users through **RQ2** survey.

Upon examining the results presented in Table 2, it is apparent that there are fewer false positives (FP) and false negatives (FN) when comparing our recommendations with developers' perceptions as opposed to retrospective evaluation. This difference can be attributed to the fact that recommending deletions involves multiple factors beyond user reviews, which Radiation does not take into account. Therefore, when asking developers to make a decision based on user reviews, Radiation demonstrates better performance. Hence, Radiation achieves an average F-score of 0.9 when its recommendations are compared with the developers' decisions based on the respective clustered reviews.

## 6.3 Results of external validation of RADIATION with users

We aim to assess the degree to which recommendations generated by Radiation align or conflict with user experience toward specific app functionalities. To answer this question, we performed a user study. In this user study, we invited the participants to our lab. The primary focus is to collect empirical data on user interactions and feedback regarding specific features, with particular attention to user experience and the potential impact of feature deletions. 42 participants took part, all aged between 18 and 32 years old. Thirty-one of them had graduated with a BSc or BA degree, while the remaining participants were pursuing their studies as undergraduate students. These participants have familiarized themselves with the app within the lab setting and at least for 20 minutes. Among these, 17 (40.4%) required more than 20 minutes to familiarize themselves with their assigned apps. Answering the questions was mandatory, and hence, the data was consistently available.

Our objective was to evaluate user sentiment towards the functionalities recommended for deletion by Radiation. To achieve this, we surveyed 42 users to evaluate their perception of specific mobile app functionalities and understand their sentiments if those functionalities were to be removed (refer to Figure 3). We asked each participant two questions regarding the features they were evaluating. Figure 5 displays a violin plot of the results. Table 3 provides a summary of the results obtained for the first survey question in RQ2, presented for each of the 25 apps under evaluation. Each column represents the average responses from three survey participants. It's important to note that the number of samples across TP (true positive), TN (true negative), and other categories varied. For instance, the app (A1) app.openconnect had only one UI functionality correctly recommended for deletion (TP) in **RQ2**, as detailed in Table 2. We also asked users how they would feel if the functionality were to be removed (Q2). We observed a high correlation of -0.86 between the responses to Q1 and Q2 in our survey. That being said, we found that the more negative the users' feelings towards the feature, the more positive they were about its removal.

When asked about the overall satisfaction for each app they were assigned, the satisfaction was rather low (with an average of 2.55 stars), with only 8.5% of the users ranking an app with five stars. The distribution of the stars is shown in Figure 5-(a). We also asked users to evaluate each feature based on the Kano schema. In the Kano schema, we asked users to evaluate each feature based on four categories: essential, worthwhile, uninteresting, or unwise. Essential features are considered necessary and form the baseline expectations, while worthwhile features add value and satisfaction. Uninteresting features don't significantly impact user satisfaction, and unwise features, if included, might even decrease satisfaction. This categorization allows us to understand how users perceive and prioritize different features, providing valuable insights into what aspects are essential or desirable for them [44, 45]. Figure 6 shows the distribution of importance among the surveyed functionalities where the wider sections of the violin indicate higher density, while narrower sections indicate lower density. The majority of features were perceived as essential or worthwhile, with only a smaller subset and a portion of participants voting for certain features as unwise.

When we asked users about the functionalities, we observed that the average sentiment of the participants towards the features that were correctly recommended for deletion by Radiation (TP recommendations) was consistently negative. In other words, the users' negative experiences were aligned with the recommendations. However, for deletions that were not actually performed (FP), we observed mixed sentiments. Nevertheless, the majority of the apps (13 out of 16) received an overall average of negative sentiments for wrong predictions as well. Thus, it is essential to note that a negative experience might not necessarily imply feature deletion but could call for a bug fix or a change in the software.

This finding aligns with our analysis of **RQ2**, where external developers favored Radiation recommendations, while historical data showed that the decisions of the actual app developers (**RQ1**) were different. This difference could be due to the exclusion of particular ecosystem or business factors in Radiation modeling. We observd that The users consistently disliked the functionalities that Radiation correctly recommended for deletion and, in general, are not against removing them.

**Table 3**: Evaluating user sentiments toward the features Radiation recommends for deletion through a survey (**RQ2**)

| App ID | Q1: Average Sentiment toward functionalities that are | | |
|---|---|---|---|
| | Incorrect deletion recom. (FP) | Correct deletion recom. (TP) | other (FN or TN) |
| (A1) | N/A | -1.3 | 2.0 |
| (A2) | -0.13 | -1.07 | 0.86 |
| (A3) | -0.7 | -1.16 | 1.0 |
| (A4) | -0.55 | -0.66 | -0.66 |
| (A5) | 1.07 | -1.0 | 0.08 |
| (A6) | N/A | -0.86 | -1.13 |
| (A7) | N/A | -0.93 | 0.0 |
| (A8) | N/A | -0.06 | 0.13 |
| (A9) | -0.66 | -1.2 | 0.91 |
| (A10) | 1.27 | -0.91 | 1.13 |
| (A11) | -0.55 | -1.0 | 1.05 |
| (A12) | -0.45 | -1.8 | -0.79 |
| (A13) | -0.56 | -1.0 | 0.51 |
| (A14) | N/A | -2.0 | 0.06 |
| (A15) | -0.88 | -1.4 | 0.79 |
| (A16) | N/A | -1.66 | 0.81 |
| (A17) | -0.77 | -1.0 | 0.73 |
| (A18) | N/A | -1.66 | 0.21 |
| (A19) | 0.97 | -1.08 | 0.91 |
| (A20) | 0 | -1.16 | 1.21 |
| (A21) | N/A | -1.13 | 0.05 |
| (A22) | -1.03 | -1.5 | -0.31 |
| (A23) | -0.89 | -1.33 | -1.09 |
| (A24) | - 0.09 | -1.55 | -0.45 |
| (A25) | -1.02 | -1.02 | 0.18 |

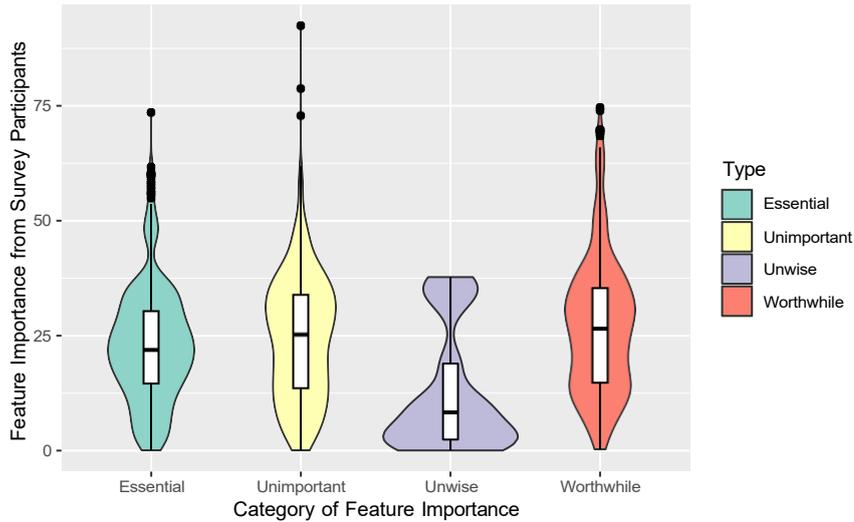

**Fig. 6**: Violin plot of feature evaluation with users based on the Kano model. The "violin" shape surrounding the boxplot displays the probability density of the data at different values.

All the features that had more than 10% unwise votes based on the Kano schema have also been positively perceived to be deleted (Figure 5). We also observed that the stronger the users voted for the importance of the feature (i.e., essential and worthwhile features), the more they feel disappointed if the feature is deleted (correlation of 0.76), which is intuitive.

# 7 Reflection: Status Que and Possibility of Incorporating Deletions in the Release Decisions

The release planning problem has traditionally been concerned with balancing conflicting priorities, such as resource constraints, time limitations, and stakeholder expectations. This intricate challenge involves making decisions about the selection and scheduling of features [49–53]for inclusion in software releases [54, 55]. The primary objective is to optimize the allocation of resources and meet strategic goals. In particular, the 'what-to-release' problem focuses on identifying features and bug reports to be added to upcoming releases to maximize the value within given constraints, both for each release and overall for the product. The next release problem is widely studied [56, 57] and, as such, revolves around the challenge of adding features and/or identifying enhancements to meet user needs and expectations in the upcoming version of the product [58–60]. As the software development landscape continues to evolve, the release planning problem remains a critical aspect of project management.

In particular, our survey focused on understanding the aspect of deletion decisions to address the planning issues and further investigate the possibility of extending the release planning definition to further include feature deletions.

Following the design science research, it is important to ground and structure the lessons learned for improving the decisions in practice [61]. As a result, we designed a survey to understand the possibility and practice of release planning in consideration of feature deletion, and based on the feedback received from the users, we performed a survey with software developers to understand their perceptions. We followed the survey design principles as outlined by Kitchenham and Pfleeger [27]. At first, we outlined three main objectives to guide our survey:

**Obj1:** Understanding frequency and cause of feature deletions.
**Obj2:** Understanding feature evaluation process and the role of users and their review in deletions.
**Obj3:** Understanding the release decisions and practices for excluding or deleting a feature.

Based on these objectives, we then move forward to designing the questionnaire and running the survey.

We took several steps to design the survey in a manner that complements the information gathered in previous works [7, 11].

## 7.1 Protocols

Initially, we initiated a brainstorming session to formulate questions addressing each objective, meticulously filtering out redundancies. Subsequently, we compiled a finalized list of 20 unique questions aligned with our objectives. To ensure the novelty of our questions and their absence in the current state of the art, we conducted a comprehensive literature review. We then curated and categorized the questions for each objective. The survey includes five demographic questions, one yes/no question, eight Likert-scale questions, and six open-ended questions. Our demographic questions address participants' software and mobile app development experience, role, size of team, and self-perceived familiarity with the process of releasing mobile apps. We designed the survey instrument using the Qualtrics platform. We ensured participant anonymity, with no collection of identifying information.

We invited five students with existing experience in app development for a pilot study to assess survey clarity, reliability, and validity. We then solicited feedback from these participants and reworded a few questions for further clarity. These responses were discarded and not used in the final analysis. We then sent the survey to our connections in the industry and advertised it on social media platforms (Twitter and LinkedIn). We received 242 clicks on the survey. Having the responses, we first checked for the completeness of responses and excluded the ones that were incomplete. We then performed descriptive statistics and visualization to report quantitative data and thematic analysis to analyze qualitative data.

We utilized our social media channels to disseminate the survey and extend invitations to developers for our study. We garnered 242 clicks, with 163 participants initiating the survey. Out of these, 141 successfully completed the survey and submitted their responses. The survey was specifically promoted for product managers and decision-makers within software teams. We used statistical and analytical techniques to analyze the numerical and categorical data. For the open text, the process was semi-manual, where two independent annotators performed sorting and aggregation [62]. In this process, one author was the moderator whenever any disagreement appeared. We also gathered the demographic information of our survey participants to contextualize the results. In our analysis, whenever appropriate, we compared the different demographic groups and reported the results.

## 7.2 Results: Developers' Perception and Implications of Deletions in Release Planning

In collecting demographic data through questions Q1 to Q5 (refer to Table 4), these developers, on average, had 8.6 years of experience in software development (ranging from a minimum of three years to a maximum of 24 years) and were involved in developing an average of 2.5 apps (ranging from a minimum of one to a maximum of nine apps). We categorized team sizes into four groups: teams with fewer than five developers, teams with five to 20 developers, teams with 20 to 50 developers, and teams with more than 50 developers. 66 participants reported working in teams of 20-50 developers, while the remaining participants (53.2%) were part of smaller teams consisting of 5-20 developers.

Table 4: Survey with developers to understand the decision process for feature deletion.

| ID | Question | Response type |
|---|---|---|
| **Demographics** | | |
| Q1 | How many years of experience do you have as a software developer? | Numerical |
| Q2 | How many mobile apps have you actively contributed to in the development process? | Numerical |
| Q3 | What is the size of your current team? | Categorical |
| Q4 | What is your current position in the team? | Short text |
| Q5 | To what extent have you been involved in the decision-making process for removing features or functionalities? | Likert scale |
| **Evaluating App Features** | | |
| Q6 | How important is it to regularly evaluate and update software features and functionality? | Likert scale |
| Q7 | How do you typically assess the impact of a deprecated feature on your existing projects? | Open text |
| Q8 | How frequently does removing a functionality change the UI elements in apps? | Likert scale |
| Q9 | How frequently are user feedback and opinions considered when deleting a functionality? | Likert scale |
| Q10 | How often do functionality deletions contribute to improving the overall user experience? | Likert scale |
| Q11 | How often do you monitor user reviews and feedback after a functionality deletion to assess the impact on user experience? | Likert scale |
| Q12 | What measures do you take to minimize negative impacts on users when functionality is deleted? | Open text |
| **Release Decisions** | | |
| Q13 | Do you conduct any specific measurements to support your decision to release a product update? | Yes/No |
| Q14 | If yes, what specific measurements do you conduct? | Open text |
| Q15 | How often do you plan for your app releases? | Likert scale |
| Q16 | How often do you plan for feature deletions? | Likert scale |
| Q17 | Who are typically involved in making decisions for feature removal? | Open text |
| Q18 | What factors typically influence your decision-making process when planning for functionality deletions? | Open text |
| Q19 | How do you balance user feedback against other factors, such as technical considerations and business requirements? | Open text |
| Q20 | How important is it to communicate the rationales behind functionality deletions, particularly in response to user reviews? | Likert scale |

| ID | Question | Distribution | High vs low # of apps | Small vs large team |
|---|---|---|---|---|
| Q6 | How important is it to regularly evaluating and updating software features? | Not at all (3), Low importance (7), Neutral (7), Important (51), Very important (73) | 0.281 | *0.031** |
| Q8 | How frequently removing a functionality changes the UI elements in apps? | Never (5), Occasionally (18), Sometimes (83), Often (29), Always (6) | 0.102 | 0.407 |
| Q9 | How frequently are user feedback and opinions considered when deleting a functionality? | Never (2), Occasionally (16), Sometimes (77), Often (35), Always (11) | 0.122 | 0.084 |
| Q10 | How often does functionality deletions contribute to improving the overall user experience? | Never (0), Occasionally (5), Sometimes (19), Often (42), Always (75) | *0.012** | 0.206 |
| Q11 | How often do you monitor user reviews and feedback after a functionality deletion to assess the impact on user experience? | Never (0), Occasionally (3), Sometimes (27), Often (61), Always (50) | 0.073 | *0.008** |

**Fig. 7**: Perception of developers on release management in consideration of functionality deletions.

These participants were product managers (41 participants), technical leads (36 participants), senior developers (22 participants), developers (21 participants), project managers (10 participants), and product owners (11 participants). When asked about the extent of their participation in release decisions, 39 participants were highly involved, 81 participants (57.4%) stated their moderate involvement, and 21 participants stated they were somewhat involved.

We conducted a survey to understand whether deletions are being planned in practice and how the decision is being made. The survey questions are detailed in Table 4. 141 developers fully answered the survey, which we characterized using the demographic questions in Section 6.3.

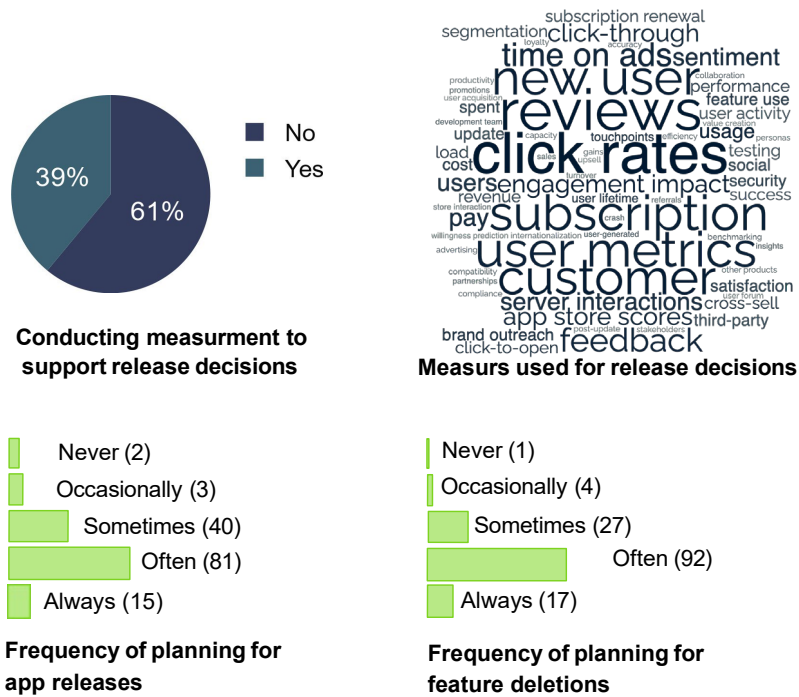

Fig. 8: The current practice of considering functionality deletion in release practices.

When asked about the importance of regularly evaluating and updating mobile app features, 87.9% of participants considered the matter important or very important. Notably, developers in larger teams (with more than 20 members) expressed a stronger emphasis on planning in this regard. Among our participants, 58.9% stated that they sometimes or somewhat frequently remove features from mobile apps, while 3.5% have never removed any features. In this decision-making process, 87.2% of participants stated that they sometimes (77 participants) or more frequently consider user feedback when deciding to exclude a functionality, while only a negligible percentage, 1.4% (two participants), have never made such a decision based on user opinions.

All participants mentioned that they occasionally or more often monitor user reviews after a feature deletion, with the majority (78.7%) doing so often (61 participants) or always (50 participants). This monitoring is more frequent and statistically more significant among teams of larger sizes ($p-value$ = 0.008). A majority of participants (53.2%) stated that such deletions always contribute to enhancing the overall user experience. However, when compared using the Mann-Whitney test, this enhancement was not significant for participants involved in developing a larger number of apps ($p-value$ = 0.012). These results are highlighted and summarized in Figure 7.

| Factors impacting feature deletion | Frequency | Participants % |
|---|---|---|
| Maintainability concerns including bugs | 91 | 64.5% |
| User feedback (positive or negative) | 86 | 61% |
| Usability challenges | 74 | 52.5% |
| Performance concerns | 51 | 36.2% |
| Adoption rates among users | 37 | 26.2% |
| Compatibility issues | 31 | 22% |
| Market trends, strategic decisions & product roadmap | 16 | 11.3% |
| Resource constraints | 7 | 4.9% |
| Regulatory compliance | 4 | 2.8% |
| Technical debt | 2 | 1.4% |

**Table 5**: Factors influencing feature removal decisions in app development. The participants were asked to name all factors in an open-text format.

When it comes to release decisions, 61% of the participants (81 participants) use measurements and analytics to inform their choices. They most frequently rely on metrics such as click rates, user metrics, user subscriptions, number of new users, and user reviews to guide these decisions. A word cloud of the responses is provided in Figure 8. The majority of these participants often plan to release their apps (57.4%). However, this number increases to 65.2% when it comes to planning in advance for excluding a feature. We summarized these results in Figure 8. Overall, developers tend to plan more diligently when deciding to delete a feature in a release compared to planning for upcoming releases in general.

In the decision-making process for feature removal, various stakeholders play distinct roles (Q17). The participants most frequently mentioned quality assurance members and testers (54.6%). This was followed by references to team leads (47.5%), product managers (45.3%), and customer relations team/manager (39%). We also asked the developers about the factors impacting their decision for feature removal in app development. The results, as summarized in Table 5, reveal that maintainability concerns, including bugs, ranked highest in frequency at 64.5%, followed by user feedback (positive or negative) at 61%. Usability challenges were noted by 52.5% of participants, while performance concerns accounted for 36.2% of responses. Other factors influencing feature deletion decisions included adoption rates among users (26.2%), compatibility issues (22%), market trends, strategic decisions, product road map considerations (11.3%), resource constraints (4.9%), regulatory compliance (2.8%), and technical debt (1.4%). Overall, User requirements stated in the form of user feedback (61%), usability concerns (52.5%), and adoption rates (26.2%) are among the most impactful criteria for planning deletions.

# 8 Discussion on Threats to Validity

Throughout the different steps of the process, there are various threats to the validity of our achieved results.

## 8.1 Construct validity - Are we measuring the right things?

We pre-processed all review texts and used machine learning classification to ensure that the analysis only considers informative user reviews. The Naive Bayes classification resulted in an F1 score of 0.82. While this is a very good result, there is still a possibility that a review has been classified incorrectly. There is a risk related to linking reviews to the proper UI elements. Two of the authors looked into the results of this linking (Step 4 of Radiation) for 600 reviews across six apps and found 71 mismatched or unrelated reviews. The influence of this noise should be considered while interpreting the results of this study. Radiation uses user reviews to recommend UI functionality deletions based on various factors. We analyzed user reviews and clustered them according to relevant UI elements, which enables Radiation to focus solely on user feedback and visible app functionality. Upon retrospective analysis, we found that Radiation has a low recall due to a considerable proportion of false negatives. These false negatives indicate deletions that were not motivated by user reviews and therefore fell outside the scope of Radiation recommendations. To further evaluate the effectiveness of our approach, we provided software developers with reviews for each UI element and asked them to decide whether they motivated functionality deletion or not. This resulted in better recall compared to our previous cross-validation results. We also evaluated user sentiment toward these functionalities and found that they consistently experienced negative emotions when using the Radiation recommended for deletion. We further discovered that the more negative the user's experience, the more likely they were to be neutral or positive about removing that feature from the app.

Additionally, we used card sorting and thematic analysis for the open-ended questions in our survey related to the survey for release decisions. The results somewhat depend on the annotators' perception and understanding of the responses. To mitigate this, annotators worked independently, adhering to a strict empirical protocol. Any disagreements were discussed and mediated to ensure consistency.

## 8.2 Conclusion validity- Are we drawing the right conclusion about treatment and outcome relation?

In comparison to studies in the context of mobile apps (Table 6), our surveys can be considered highly participated. However, we used non-compensated and convenience sampling to attract participants, which might bias the conclusions that are drawn [63]. It is essential to note this type of evaluation is subjective. However, in RQ1, the results of the retrospective analysis of the data are aligned with our survey results with developers (external validity) and the users' perception (**RQ2**). In total, we think that the evaluation gained with 37 developers and 42 users is sufficient to confirm our findings.

When connecting a review to a UI element in Radiation, there is a chance that we relate a review to an element incorrectly (false positives). This may happen because

- We may miss some UI elements, as they can be instantiated in the program code or hard coded,
- Some UI elements are not visible to the end user, or
- Text of some UI elements are common English words or can have similar labels in different app views.

To address the first two items above, we used Backstage [26] on a few of the apps, and we found that while the risk exists, it is relatively small. Since Backstage works on compiled application binaries we were limited to using it in Radiation. For the third item above, we applied pre-processing as suggested in CRISTAL [18] and adopted their list of stop words. Further, Radiation is not intended to find all the deleted features (recall) exhaustively. The impact of potentially missed elements is insignificant.

We validated our approach in two stages: first, by looking back at past decisions made by app developers and comparing them with recommendations from Radiation. Then, in the second stage, we shifted our focus to see if human experts, using the same data, would come to similar conclusions as Radiation. In this situation, we sought the advice of experienced software developers who were not involved in creating the apps we were studying. This helped us understand how meaningful analogical reasoning is in this context. Analogical reasoning is identifying similarities between two different concepts or situations and using this comparison to gain insights or solve problems. In our study, we refer to analogical reasoning as the method by which users draw parallels between familiar elements and new functionalities to make decisions or provide feedback. This concept is closely related to case-based reasoning (CBR), where users apply knowledge from previous cases to new situations. Both analogical reasoning and CBR involve leveraging past experiences to evaluate and understand new information. By defining and exploring analogical reasoning, we can better interpret user feedback and improve the design and functionality of user interface elements. This connection allows us to develop more intuitive and user-friendly interfaces by understanding how users transfer knowledge from known contexts to new ones [64]. We made this choice for two main reasons: firstly, because we didn't receive responses from the original developers, and secondly, to add depth to our evaluation of the research question. It's important for readers to keep in mind this limitation when they're interpreting the results.

Another potential threat arises from the level of familiarity users have with the mobile apps during evaluation in RQ2. All participants in the study interacted with the application for a minimum of 20 minutes in controlled laboratory conditions. This may raise concerns regarding the conclusion's validity. However, we argue that the controlled setting in the laboratory and the dedicated time allotted for app usage mitigate the typical randomness inherent in surveys, instilling a higher degree of confidence in the results. Without this designated time in the lab, there is no guarantee that participants have actually used the app or are up to date with the recent functionality, potentially leading to random responses. Conversely, obtaining contact information and accessing individuals who have independently used these applications, especially at scale, is not feasible. Thus, we conducted the lab study to ensure controlled usage conditions.

### 8.3 Internal Validity - Can we be sure that the treatment indeed caused the outcome?

The selection of attributes used in Radiation to decide *if a UI functionality should be deleted* is another threat to validity. Our survey with users was aligned with the findings in the literature [7] and showed that users and their feedback are important information in the deletion process. However, it is not the only decisive factor for excluding a functionality from apps. We selected attributes based on related studies (Table 1). There are other attributes related to competitors, performance, or maintenance considerations that are relevant for the decision-making but could not be taken into account for our study. Following the results of former studies on mobile apps [18], we assumed that users are reviewing just the functionality that is visible to them (and not the background code). This might not be true for all the users, reviews, and sentiments. However, we expect a low number of such cases.

### 8.4 External Validity - Can the results be generalized beyond the scope of this study?

Our retrospective analysis was performed on open-source mobile apps. The number of apps, reviews, and commits analyzed is considered high, indicating that results are significant, at least for open-source mobile apps. While selecting the apps for this study, we did not consider their status (for example, the number of downloads), which may pose a risk of bias in the findings. The results may vary between apps with regard to their status on the app store. Also, we have not discussed apps outside Google Play, such as iTunes. Hence, the results might not be representative of all apps. However, the choice of sample size and the platform is comparable with state-of-the-art studies [12].

When it comes to surveys with developers and user studies, our study draws conclusions based on a survey, which can be inaccurate at times. Surveying software developers does not always provide a comprehensive perspective of real-world practices [65]. We used surveys to triangulate the results of our internal validation (**RQ1**) and to gain a deeper understanding of practices related to deletion decisions.

## 9 Related Work

In this study, we challenged Lehman's law of growth by investigating functionality deletion as a specific activity in the development process. We focused on the mobile apps because the device resources are limited and the size of the release has been introduced as a decisive factor for release decisions [66]. Feature and functionality deletion for software products in general have been discussed mostly on the model level, which triggered us to widely investigate on the nature and reasons for functionality deletion in **RQ1**. Development activities in software engineering involve adding, deleting, and modifying elements [67, 68]. However, discussions have primarily focused on adding and modifying features, with less attention given to deletion. Adding new functionality is a key consideration in release planning, and existing approaches often concentrate on this aspect [4, 5, 69], or they revolve around handling change requests [29, 70]. Commonly in literature, studies have used the term "code churn", which represents the total count of added or deleted lines of code. However, this metric does not differentiate between additions and deletions [71, 72]. Murphy-Hill et al. [73] made an important distinction between adding and deleting features, particularly in the context of bug fixes. They defined *functionality deletion* as the removal of a feature during a bug fix. Their findings indicate that 75% of participating developers remove functionality to address bugs.

Furthermore, the release planning of mobile apps and/or the analysis of user reviews to support app evolution and maintenance have been studied by several researchers [12? ]. In the following sections, we discuss the state of the art in these studies as they pertain to requirements engineering, specifically addressing RQ1 and RQ2. Our primary focus is to explore the potential and feasibility of recommending feature deletions rather than striving for perfect accuracy in these suggestions. Therefore, we rely exclusively on the best practices documented in the current literature without engaging in benchmarking or method comparisons. Additionally, we present a summary of related work on release planning, focusing on survey analysis in particular.

## 9.1 User Reviews to Support Apps' Evolution

Table 6: Context and evaluation of related studies.

| Method | Context | Evaluation |
|---|---|---|
| ARdoc [33] | Information giving/seeking, feature request, problem discovery, others | Evaluating three apps by two developers |
| AR-Miner [31] | Informative or non-informative reviews | Manual inspection by authors, comparison between techniques |
| ChangeAdvi [37] | Localizing change request by linking reviews to the source code | Evaluated results with 12 developers |
| CLAP [29] | New feature request, bug report | Retrospective analysis of 463 reviews and interview with three developers |
| CRISTAL [18] | Tracing user reviews to the developers changes | Manual evaluation by authors |
| MARA [74] | Feature request | Comparing different techniques |
| PAID [75] | Issues (bugs) | Retrospective analysis of 18 apps |
| Panichella et al. [20] | Information giving/seeking, feature request, problem discovery, others | Comparison between different methods |
| SURF [14, 17] | Information giving/seeking, feature request, problem discovery, others | 23 developers analyzed SURF output for 2622 reviews. |
| SUR-Miner [30] | Aspect evaluation, praise, function request, bug report, others | Comparing different techniques, evaluation with 32 developers |
| URR [21] | Compatibility, usage, resources, pricing, protection, complaint | Qualitative evaluation by a student and a developer |

Analyzing user reviews to support app evolution and maintenance has been explored by various researchers [12]. These studies primarily focus on differentiating user needs, categorized as either "feature requests" or "bug reports" [32]. Notably, Palomba et al.'s study [18] revealed that 49% of informative reviews were considered for app evolution. In these investigations, user reviews serve as sources of change requests, employing various Natural Language Processing (NLP) techniques to offer prioritization or classification schemes. The goal is to assist developers in deciding on the next best changes, whether by adding new functionality or addressing a bug. Table 6 provides an overview of the most relevant methods.

CLAP [29] adopted a mixed method, combining retrospective analysis of changes for 463 reviews with interviews of three app developers. PAID [75] conducted a comprehensive retrospective evaluation by investigating 18 apps for issue prioritization. In comparison, our study involved a more rigorous evaluation, with 37 developers assessing 36,039 reviews across 25 apps. We further compared these evaluations with the results obtained from Radiation.

While some studies compared different evaluation methods, this was not feasible for Radiation in general, as none of the existing techniques focused on functionality deletion. However, for selecting classifier and topic modeling techniques, we made the comparisons, as discussed in Section 4.

## 9.2 Release Planning and Mobile Apps

Release planning is a well-established practice in software requirements engineering [54, 76]. Numerous studies have explored various aspects of this critical phase, delving into prioritizing features, resource allocation, and decision-making processes within release planning. Release planning is often being addressed as a search problem [77]. These investigations contribute valuable insights to enhance the effectiveness and efficiency of software release planning strategies. To gain a comprehensive understanding of the overall state of release planning, we leveraged existing systematic literature reviews. Achimugu et al. [78] identified 73 papers in the context of requirement prioritization and planning in their systematic literature study, while Riegel and Doerr [79] analyzed 83 papers. Notably, these two studies had 15 papers in common, and we carefully reviewed them by inspecting the abstracts. However, none of these papers mentioned or considered feature deletions. Similarly, the studies gathered by Svahnberg et al. [69] also did not address the deletion of features.

Similarly, starting from the existing systematic literature review [12, 80, 81], numerous studies in the realm of mobile apps have addressed the planning and prioritization of requirements for decisions regarding additions to the next release. However, none have delved into the discussion of the deletion of app features. Our search within the recent body of literature has yielded no papers considering the deletion of mobile apps other than those from our own research, which we elaborate on in this study. However, a number of studies have discussed the requirements prioritization along with the need for release planning [29, 66, 82–84]. The current literature primarily addresses different user requests in app evolution, while our study focuses on a functionality deletion — and its triggers.

## 10 Future Work

Overall, the main goal of future research will be to better understand the deletion of functionality as part of software evolution, also beyond mobile apps. One key motivation for the paper comes from the observation that current release planning in general [5] and in particular for mobile apps [12, 29] is exclusively focused on feature addition. Planning in consideration of both addition and deletion of functionality requires revisiting the planning objective(s). Clearly, deletion consumes development effort as well. While we took the first step toward understanding functionality deletion, future work involves contextualizing the results for specific projects and development teams. Besides a more comprehensive empirical evaluation in general, we also target trade-off analysis between measuring the evolving maintenance effort and functionality deletions [85–89].

In addition, we will work on improving the performance of our recommendations by updating the machine learning techniques and features and tuning the model (for instance, by more in-depth analysis of similarity). We relied on the highly performed methods discussed in the literature and did not re-evaluate the performance of the learners. We do not argue these techniques are the most optimal and highest-performing methods possible. Rather, as the first study on recommending feature deletion in app releases, we focused on exploring the possibility of deletion recommendations, their usefulness, and the ease of explanation to the users and the developers.

As the first study on predicting deletions based on user reviews, our target was to examine if the deletion prediction is possible rather than to highly optimize the performance of the approach. This is an essential step before taking further steps to plan these deletions. Based on the current state-of-the-art results, we do not expect that a benchmark of different classifiers would significantly improve the performance of our approach. The results of our survey with practitioners show a systematic approach toward planning deletions in mobile apps, combined with the measurement of a variety of factors (see Figure 8 and Table 5). This discussion explores the possibility and potential of planning for deletions in the software development life cycle. Recognizing the importance of systematically considering the removal of features adds a new dimension to release planning, emphasizing a holistic approach that encompasses both additions and deletions to enhance overall product development and maintenance strategies.

## 11 Conclusions

*Lehman's law on continuous growth of functionality does not universally apply.* In the domain of mobile apps, developers frequently delete functionality—be it to fix bugs, maintain compatibility, or improve the user experience. We performed a study with *app users* to confirm the potential value of deletions also from their perspective. We suggested that the process of selecting the functionality to be deleted can be automated, as demonstrated by our Radiation recommendation system. Radiation analyses the UI elements of the app and the reviews and recommends if the UI element and its functionality shall be deleted or not. We further conducted a study with users to understand their perception of the features recommended for deletion by our method. Additionally, our survey with developers revealed that they carefully plan when deciding to remove a feature, with user and usage data playing a crucial role in these decisions. This is the first study to investigate the prediction of functionality deletion in software evolution. It opens the door towards a better understanding of software evolution, in particular in an important domain such as mobile app development. In the days of Lehman's studies, features such as user experience, screen space, or energy consumption were not as crucial as they are today; it may be time to revisit and refine Lehman's findings.

## 12 Data and Code Availability Statement

DATA: The related artifacts of this paper are available at https://github.com/maleknaz/Radiation. The data of mobile apps are subjected to Google Play copyright, and hence, we cannot openly provide access to them. Our dataset is hosted on GitHub to ensure maintainability and ease of updates while adhering to the legal terms applicable to data hosted on mobile app marketplaces. The data was collected exclusively for this study, with no commercial or proprietary use intended, and has been managed in accordance with the relevant terms and conditions. To request access to the dataset, please contact us directly. Each request will be evaluated individually to ensure full compliance with all legal requirements.

CODE: We used three primary tools and their associated code in the RADIATION steps S1 to S7 [19, 26, 31], as referenced throughout the paper. We foresee future research investing in the improved performance of these technologies.